\documentclass[%
reprint,
superscriptaddress,
showpacs,preprintnumbers,
notitlepage,
 amsmath,amssymb,
aps,
prx,
floatfix
]{revtex4-1}

\usepackage{float}
\usepackage{graphicx}
\usepackage{dcolumn}
\usepackage{bm}
\usepackage[multidot]{grffile}
\usepackage[colorlinks=true,allcolors=blue, breaklinks=true]{hyperref}
\newcommand{\vect}[1]{\bm{\mathrm{#1}}}

\newcommand{\lap}{\vect{L}}
\DeclareMathOperator{\ima}{Im} 

\usepackage[dvipsnames]{xcolor}

\usepackage{xspace}

\usepackage[utf8]{inputenc}

\newcommand{\new}[1]{\textcolor{black}{#1}}
\newcommand{\neww}[1]{\textcolor{black}{#1}}
\begin{document}


\title{\new{Multiorder Laplacian for synchronization in higher-order networks}}

\author{Maxime Lucas}
\email{maxime.lucas.1@univ-amu.fr}
\affiliation{Aix Marseille Univ, CNRS, CPT, Turing Center for Living Systems, Marseille France}
\affiliation{Aix Marseille Univ, CNRS, IBDM, Turing Center for Living Systems, Marseille France}
\address{Aix Marseille Univ, CNRS, Centrale Marseille, I2M, Turing Center for Living Systems, Marseille France}

\author{Giulia Cencetti}
\affiliation{Mobs Lab, Fondazione Bruno Kessler, Via Sommarive 18, 38123, Povo, TN, Italy}

\author{Federico Battiston}
\email{battistonf@ceu.edu}
\affiliation{Department of Network and Data Science, Central European University, Budapest 1051, Hungary}

\date{\today}
\begin{abstract}
The emergence of synchronization in systems of coupled agents is a pivotal phenomenon in physics, biology, computer science, and neuroscience. Traditionally, interaction systems have been described as networks, where links encode information only on the pairwise influences among the nodes. Yet, in many systems, interactions among the units take place in larger groups. Recent work has shown that the presence of higher-order interactions between oscillators can significantly affect the emerging dynamics. However, these early studies have mostly considered interactions up to 4 oscillators at time, \new{and analytical treatments are limited to the all-to-all setting.} Here, we propose a general framework that allows us to effectively study populations of oscillators where higher-order interactions of all possible orders are considered, \new{for any complex topology described by arbitrary hypergraphs, and for general coupling functions.} To this end, we introduce a multiorder Laplacian whose spectrum determines the stability of the synchronized solution. Our framework is validated on three structures of interactions of increasing complexity. First, we study a population with all-to-all interactions at all orders, for which we can derive in a full analytical manner the Lyapunov exponents of the system, and for which we investigate the effect of including attractive and repulsive interactions. Second, we apply the multiorder Laplacian framework to synchronization on a synthetic model with heterogeneous higher-order interactions. Finally, we compare the dynamics of coupled oscillators with higher-order and pairwise couplings only, for a real dataset describing the macaque brain connectome, highlighting the importance of faithfully representing the complexity of interactions in real-world systems. Taken together, \new{our multiorder Laplacian allows us to obtain a complete analytical characterization of the stability of synchrony in arbitrary higher-order networks, paving the way towards a general treatment of dynamical processes beyond pairwise interactions.}
\end{abstract}

\maketitle





\section{Introduction} \label{sec:introduction}

The emergence of order in populations of interacting oscillators -- a phenomenon known as synchronization -- is ubiquitous in natural and man-made systems~\cite{pikovsky2003synchronization, strogatz2004sync}. Typical examples of synchronization include the flashing of fireflies, or the clapping of an audience. In the last decades, synchronization has been the subject of intense research, and it has been applied to a wide range of areas, including neuroscience~\cite{cumin2007generalising}, circadian rhythms~\cite{leloup1999limit}, or the cardio-vascular system~\cite{lotrivc2000synchronization, suprunenko2013chronotaxic}. In particular, much attention has been devoted to unveiling the relationship between the structure of the network of interactions and the emerging collective behavior~\cite{arenas2008synchronization, rodrigues2016kuramoto}. As an outcome of these investigations, noticeable examples include the discovery of abrupt synchronization induced by degree-frequency correlation~\cite{gomez2011explosive}, and cluster synchronization induced by structural symmetries~\cite{pecora2014cluster}. 

Most interacting systems have so far been represented as networks, a collection of nodes and links describing relationship and influences between them at the level of pairs. However, many real-world systems are better modeled by including higher-order interactions, i.e. interactions between more than two nodes at a time~\cite{battiston2020networks}. A typical example is that of human collaborations, which often occur at the level of groups. In this case, a traditional network representation is misleading, as it would associate the same structure -- a triangle -- both with the case of a triplet of people collaborating on a single task, and with the case of three individuals collaborating as three distinct pairs in different projects. This indistinguishability can be solved by making use of higher-order network representations, such as hypergraphs~\cite{berge1984hypergraphs} and simplicial complexes~\cite{aleksandrov1998combinatorial}. A stream of research has recently focused on correctly characterizing the structure of systems with higher-order interactions~\cite{courtney2016generalized, linial2006homological, kahle2014topology, young2017construction, ghoshal2009random, wang2010evolving, wu2015emergent}. Interestingly, considering this additional level of complexity sometimes leads to changes in the emerging dynamics of a complex system, including social contagion~\cite{iacopini2019simplicial, dearruda2020}, activity driven models~\cite{petri2018simplicial}, diffusion~\cite{torres2020simplicial,schaub2018random}, random walk~\cite{zhou2007learning,carletti2020random} and evolutionary games~\cite{alvarez2020evolutionary}.

Higher-order interactions can influence the nature of the dynamics also for systems of coupled oscillators. A few studies have recently investigated their effect  experimentally~\cite{jia2015experimental,matheny2019exotic}, and from a network inference point of view~\cite{rosenblum2007self,kralemann2014reconstructing,rosenblum2019numerical}. From a theoretical point of view, higher-order interactions were considered in the context of global nonlinear coupling~\cite{rosenblum2007self} and \new{multi-population resonance~\cite{komarov2013dynamics}}, they were shown to arise from phase reduction beyond the first approximation~\cite{bick2016chaos,bick2016chaotic,ashwin2016identical,leon2019phase, matheny2019exotic}, and they can facilitate chaotic behavior~\cite{bick2016chaos} \new{and other exotic dynamical regimes~\cite{matheny2019exotic}}. The Kuramoto model, where phase oscillators interact in pairs, is often invoked as the most simple way to describe the emergence of synchronization in a population of interacting nodes with local dynamics. \new{Only a few works have so far considered higher-order generalizations~\cite{tanaka2011multistable,komarov2015finitesizeinduced,skardal2019abrupt,skardal2019higher},
showing the promotion of cluster synchronization~\cite{tanaka2011multistable,komarov2015finitesizeinduced, skardal2019abrupt} and explosive transitions~\cite{skardal2019abrupt,skardal2019higher}. Yet, for all these studies, analytical insights are limited to all-to-all coupling settings, disregarding the rich architecture of interactions of real-world systems. Interestingly, a different type of model was introduced in~\cite{millan2019explosive}, where a Kuramoto phase oscillator is associated with each simplex.}


\new{In this work, we provide a full analytical treatment for synchronization in  a population of coupled oscillators where arbitrary higher-order interactions of all orders are possible.} To this end, we study a generalization of the Kuramoto model of identical phase oscillators to \new{general group interactions. These can be conveniently described by hypergraphs, in the most flexible mathematical representations of higher-order interactions, which also generalizes other commonly used formalisms, such as simplicial complexes}. For this model, we show that the stability of the fully synchronized state is determined by the eigenvalues of a newly defined \textit{multiorder Laplacian}, which takes into account the higher-order complex topology of interactions. We validate this Laplacian framework on several toy-models describing higher-order interactions of increasing complexity. First, we investigate all-to-all interactions at all orders, for which the eigenvalues can be derived fully analytically. We further characterize this system by considering three subcases: (i) attractive coupling only, (ii) interplay between attractive and repulsive orders, (iii) and decaying coupling strength (that we link to higher-order phase reduction studies). We confirm our analytical findings with numerical simulations. Second, we consider the star-clique model, a toy model specifically generated to highlight some simple spectral properties of the multiorder Laplacian. Finally, we investigate the effect of higher-order interactions on synchronization on a real macaque brain dataset. Taken together, our work sheds new light on the effect of higher-order interactions in a population of coupled oscillators, and it unveils how new emergent phenomena can be captured analytically through the introduction of a suitable Laplacian framework, which naturally generalizes the traditional approach to networks beyond pairwise interactions.


\begin{figure*}[thb]
	\centering
	\includegraphics[width=0.9\linewidth]{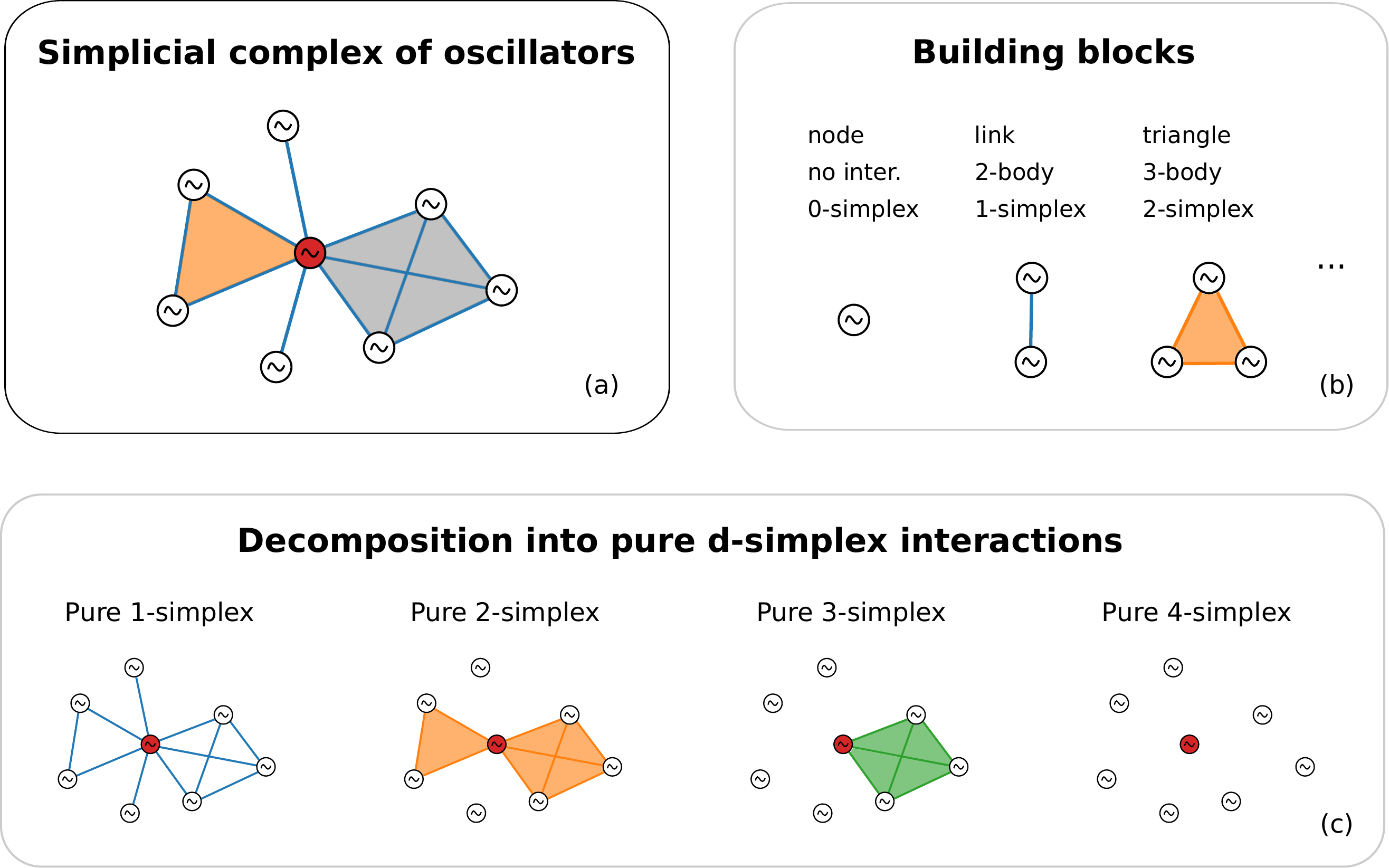}	
	\caption{{\bf Population of oscillators with higher-order interactions on simplicial complexes. } (a) Example of simplicial complex of oscillators: the red node has higher-order interactions of orders up to 3 (4-oscillator) with the other oscillators. (b) Here, the building blocks of the higher-order interactions consist in edges (1-simplices), triangles (2-simplices) and tetrahedra (3-simplices). A $d$-simplex represents a $(d+1)$-oscillator interaction. (c) The simplicial complex (a) can be decomposed into its pure $d$-simplex interactions: the red node has seven 1-simplex interactions (blue), four 2-simplex interactions (orange), and one 3-simplex interaction (green), but no 4- or more simplex interactions. In (a), the faces of the tetrahedra belong to 2-simplices (orange) and to a 3-simplex (green) so that they are depicted in gray. 
	}
	\label{fig:fig1simplicialnet}
\end{figure*}

\section{Generalized higher-order interactions} \label{sec:system}

We study the effect of higher-order interactions in a population of $N$ identical phase oscillators. Specifically, we consider the dynamics of oscillators with the most general topology describing many-body interactions of any order $d=1,\ldots,D$ 
%
	\begin{equation}
	\begin{aligned}
	\dot \theta_i = & \,\, \omega + \frac{\gamma_1}{\langle K^{(1)} \rangle} \sum_{j=1}^N A_{ij} \sin(\theta_j - \theta_i) \\
	& + \frac{\gamma_2}{2! \langle K^{(2)} \rangle} \sum_{j,k=1}^N B_{ijk} \sin(\theta_j + \theta_k - 2 \theta_i) \\ 
	& + \frac{\gamma_3}{3! \langle K^{(3)} \rangle} \sum_{j,k,l=1}^N C_{ijkl}  \,  \sin (\theta_j + \theta_k + \theta_l - 3 \theta_i) \\ 
	& + ... \\
	& + \frac{\gamma_{D}}{{D}! \langle K^{({D})} \rangle} \sum_{j_1, \ldots, j_{{D}}=1}^N \, \, \,  M_{ij_1 \ldots j_{D}} \sin \left( \sum_{m = 1}^{{D}} \theta_{j_m} -  {D} \, \theta_i \right) ,
	\end{aligned}
	\label{eq:system}
	\end{equation}
%
a natural generalization of the Kuramoto model, where $\omega$ is the natural frequency of each oscillator, $\gamma_1, \gamma_2, \ldots, \gamma_D$ are the coupling strengths at each order, the \neww{$\langle K^{({1})} \rangle, \dots, \langle K^{({D})} \rangle$ are the average degrees at order $1, \dots, D$ (explicitly defined in Eq.~\eqref{eq:d-degree})}, and the adjacency tensors $\vect{M}$ determine the topology. Just like $A_{ij}=1$ if there is a pairwise interaction $(i,j)$ but 0 otherwise, $B_{ijk}=1$ if there is a triplet interaction $(i,j,k)$ but 0 otherwise, and similarly for all orders. Note that the interactions are assumed undirected, i.e. the adjacency tensors are invariant under any permutation of their indices. These adjacency tensors encode the most general topology of higher-order interactions that can be formalized as hypergraphs or simplicial complexes, for example. The largest value that $D$ can take is $N-1$, which corresponds to $N$-oscillator interactions, the highest order possible. \new{The general interaction scheme of (1) is illustrated with an example in Fig.~\ref{fig:fig1simplicialnet}. There, a 3-oscillator interaction is represented by a 2-simplex, and any $(d + 1)$-oscillator interaction is represented by a $d$-simplex, (also called a simplex of order d), as illustrated in Fig.~\ref{fig:fig1simplicialnet}(b). Fig.~\eqref{fig:fig1simplicialnet}(c) shows a visualization of each pure order $d$.} 

The sinusoidal coupling functions are chosen as a natural generalization of those used in the Kuramoto model: when describing the dynamics of oscillator $i$, they are symmetric with respect to $i$, meaning that any permutation of the other indices leaves them invariant. \new{Even when restricting ourselves to $2\pi$-periodic functions that vanish when oscillators are identical, there exists more choices at larger orders. At order 2, for example, the only other choice is $\sin(2\theta_j - \theta_k - \theta_i)$, used e.g. in~\cite{skardal2019higher}. All these other choices are discussed in App.~\ref{app:coupling_functions} and do not affect the generality of our framework. For the sake of clarity of our presentation, however, we do not include them earlier. } \neww{Sinusoidal couplings naturally arise when applying phase reduction to more realistic nonlinear models, for the Kuramoto model at first order, but also at higher orders~\cite{bick2016chaos,leon2019phase}, e.g. for nanoelectromechanical oscillators~\cite{matheny2019exotic}.}

\new{A few studies have investigated systems similar to~\eqref{eq:system} analytically, but only in all-to-all schemes~\cite{tanaka2011multistable,komarov2015finitesizeinduced,skardal2019abrupt,skardal2019higher,gong2019lowdimensional}. Insights into complex topologies, for instance the onset of synchronization~\cite{skardal2019higher}, have so far been limited to numerical simulations, and a formal description of the process is still an open problem. In Sec.~\ref{sec:laplacian}, we introduce an analytical framework that allows us to overcome current limitations and investigate arbitrarily complex topologies. Such general patterns are formalized as hypergraphs, which are the mathematical structures that allow for the most general encoding of higher-order interactions. 
These include -- but are not limited to -- simplicial complexes, more constrained higher-order representations where a group interaction of order $d$ requires all its lower order subgroup interactions, illustrated in Fig.~\ref{fig:fig1simplicialnet} for model~\eqref{eq:system}.} 

For the introduced model~\eqref{eq:system}, the existence of the fully synchronized state, $\theta_i=\theta_j$ for all $i$ and $j$, is trivially guaranteed and implies the solution $\theta_i(t) = \omega t$. In this paper, we focus on the stability of that fully synchronized state. For convenience, we start by going to the rotating reference frame $\psi_i = \theta_i - \omega t$. This is equivalent to applying the transformation $\theta_i \mapsto \psi_i$ and $\omega \mapsto 0$ to the original system~\eqref{eq:system}. In this new reference frame, the synchronized solution is given by $\psi_i(t) = 0$ for all $i=1,...,N$. 

The linear stability of the synchronized state is determined by the dynamics of heterogeneous perturbations $\delta \psi_i(t)$, which satisfy the linearized dynamics
%
	\begin{equation}
	\begin{aligned}
	\delta \dot  \psi_i = &+ \frac{\gamma_1}{\langle K^{(1)} \rangle} \sum_{j=1}^N A_{ij} (\delta \psi_j - \delta \psi_i) \\
	& + \frac{\gamma_2}{2! \langle K^{(2)} \rangle} \sum_{j,k=1}^N  B_{ijk} (\delta \psi_j + \delta \psi_k - 2 \delta \psi_i) \\ 
	& + \frac{\gamma_3}{3! \langle K^{(3)} \rangle} \sum_{j,k,l=1}^N  \,  C_{ijkl} \, (\delta \psi_j + \delta \psi_k + \delta \psi_l - 3 \delta \psi_i) \\ 
	& + ... \\
	& + \frac{\gamma_{D}}{{D}! \langle K^{({D})} \rangle} \sum_{j_1, \ldots, j_{{D}}=1}^N \, \, \,  M_{ij_1 \ldots j_{D}} \left( \sum_{m = 1}^{{D}} \delta \psi_{j_m} -  {D} \, \delta \psi_i \right) .
	\end{aligned}
	\label{eq:linearised_system}
	\end{equation}
%
\neww{These equations are straightforwardly obtained by linearizing system~\eqref{eq:system}. It is worth noticing, however, that the same equations could also be derived by assuming a different (not necessarily sinusoidal) shape of the coupling functions in the original system, via the linearization mechanism.}
For networks with pairwise interactions only, the dynamics of those perturbations -- and hence the stability of the system -- is typically assessed by using the so-called Laplacian formalism. In the next section, we see how we can characterize the stability of system~\eqref{eq:system} with higher-order interactions up to any order $d$ by extending the traditional Laplacian formalism to systems with any type of higher-order interactions. 

%
%
%


\section{multiorder Laplacian} \label{sec:laplacian}

Different generalizations of the Laplacian operator have been proposed so far in the literature to include higher orders of interactions: from the simplest versions for uniform hypergraphs \cite{chung1993laplacian, lu2011high}, to those more complicated associated to simplicial complexes~\cite{muhammad2006control, rosenthal2014simplicial, parzanchevski2017simplicial} and Hodge Laplacians~\cite{lim2015hodge, schaub2018random}, to mention a few. Let us notice that these Laplacians describe the hierarchy among building blocks of the topology, and different orders are associated to Laplacian matrices of different sizes, where the order zero is the traditional node point of view; the first order represents the edge perspective where the Laplacian size is equal to the number of pairwise connections and each entry is associated to edge adjacency; the second order Laplacian has a different  size again, being based on the existing triangles, and so on.

Here, we propose a multiorder Laplacian: an operator that generalizes the typical pairwise Laplacian framework and allows us to analytically describe the effect of higher-order couplings on node oscillatory dynamics for any simplicial interactions. This is different from the previously defined Laplacians, where interactions between, e.g. triangles, are seen from the point of view of triangles and not from the point of view of the nodes in those triangles. 

First, we show that each $d$-simplex interaction term in Eq.~\eqref{eq:linearised_system} can be written in terms of a generalized Laplacian of order $d$. Second, we display how the full system~\eqref{eq:linearised_system} can be written in terms of a multiorder Laplacian. 

\subsection{Laplacian}
\label{subsec:laplacian}

We introduce a generalized Laplacian of order $d$
\begin{equation}
L^{(d)}_{ij} =  d K^{(d)}_i \delta_{ij}  -  A^{(d)}_{ij}  ,
\label{eq:d-laplacian}
\end{equation}
\new{with the Kronecker delta $\delta_{ij}$}, and where we have defined at order $d$, the degree $K^{(d)}_{i}$, i.e. the number of distinct $d$-simplices node $i$ is part of, and the adjacency matrix $A^{(d)}_{ij}$, i.e. the number of distinct $d$-simplices the pair of nodes $(i,j)$ is part of,
\begin{align}
K^{(d)}_{i} &= \frac{1}{d!} \sum_{j_1, \ldots, j_{{D}}=1}^N \, \, \,  M_{ij_1 \ldots j_{D}} , \label{eq:d-degree} \\
A^{(d)}_{ij}& = \frac{1}{(d-1)!} \sum_{j_2, \ldots, j_{{D}}=1}^N \, \, \,  M_{ij_1 \ldots j_{D}} . \label{eq:d-adj}
\end{align}
Note that these definitions are natural generalizations of their pairwise counterparts to which they reduce when $d=1$. This newly defined Laplacian can be shown to have the expected properties of  a standard Laplacian matrix: it is symmetric, and its rows sum to zero. Moreover, its eigenvalues are all non-negative, as we shall see in the next section.

With those quantities, each term of the linearized equation \eqref{eq:linearised_system} can be rewritten as
\begin{equation}
\delta \dot \psi_i = - \frac{\gamma_d}{\langle K^{(d)} \rangle} \sum_{j=1}^N L^{(d)}_{ij} \delta \psi_j .
\label{eq:d-pert}
\end{equation}
\new{as shown in detail in App.~\ref{app:laplacian}}. We now have all the ingredients to treat oscillators at each order of interaction and build a multiorder Laplacian.


\textbf{multiorder interactions.} We go back to our original system~\eqref{eq:system}, \new{with interactions at orders $d=1,\dots,D$ combined}. We know that the stability of the synchronized solution of system~\eqref{eq:system} is determined by system~\eqref{eq:linearised_system}, which we now can write
\begin{equation}
\delta \dot \psi_i =  - \sum_{j=1}^N L^{\text{(mul)}}_{ij} \delta \psi_j ,
\label{eq:systemLapl}
\end{equation}
where we have defined
\begin{equation}
L^{\text{(mul)}}_{ij} = \sum_{d=1}^{\new{D}} \frac{\gamma_d}{\langle K^{(d)} \rangle}  L^{(d)}_{ij} ,
\label{eq:laplacian_tot}
\end{equation}
the multiorder Laplacian $L^{\text{(mul)}}_{ij}$ as a weighted sum of the Laplacian matrices of order $d$. The weight given to each order is proportional to $\gamma_d$, and normalised by the average degree of order $d$. Hence, by definition, the multiorder Laplacian gives an equal weight to each order, even if the network contains more, say, 2-simplices than 5-simplices. Notice that this normalization is not included in the definition of Laplacians of pure order $d$. \new{This newly defined Laplacian reduces to the usual operator when $D=1$, i.e. when only pairwise interactions are taken into account. Finally, we note that $L^{\text{(mul)}}_{ij}$ depends on $D$, whose maximum value is limited by the network size, since when $D=N-1$, all possible orders are considered.}

In this section, our generalized framework showed us two things. First, how to rewrite interactions at each order with a Laplacian matrix $L^{(d)}_{ij}$ of order $d$. And second, how to rewrite the full system, including all higher-order interactions, with a multiorder Laplacian matrix $L^{\text{(mul)}}$. Additionally we showed how the latter matrix is just a weighted sum of the former matrices. 

So far, we have an analytical expression for the Laplacian matrix of a given simplicial complex. It is the eigenvalues of this Laplacian that quantify the stability or instability of the synchronized state of the system. \new{We note that, even though the multiorder Laplacian is the weighted sum of each Laplacian of order $d$, its eigenvalues cannot be obtained in general as a linear combination of the Laplacians at each order $d$, as the eigenvalue operator is nonlinear}. In general, we need to numerically compute the eigenvalues of the Laplacian. While this is generally true, special cases exist where the eigenvalues can nonetheless be summed and the system characterized in a fully analytical manner. We present such case in Sec.~\ref{sec:all-to-all}.


Before this, let us make a short didactic digression and step back to analyze the pure order 2. We show the details of the Laplacian derivation in this specific case and leave those for order 3 and $d$ in the Appendix \ref{app:laplacian}.

\textbf{Pure 2-simplex interactions.}  We now rewrite the 3-oscillator interaction term in Eq.~\eqref{eq:linearised_system}, i.e. those interactions represented by filled orange triangles in Fig.~\ref{fig:fig1simplicialnet}, with our Laplacian framework. \new{The first simple mathematical step represents an important point of the derivation, allowing us to write the 2-simplex interaction of 3 phases as 2 identical terms of the difference of only 2 phases:}
\begin{align}
\delta \dot \psi_i = & \frac{\gamma_2}{2! \langle K^{(2)} \rangle} \sum_{j,k=1}^N B_{ijk} ( \delta \psi_j + \delta \psi_k - 2 \delta \psi_i)  \label{eq:2-simplex} , \\
=& \frac{\gamma_2}{\langle K^{(2)} \rangle} \sum_{j,k=1}^N  B_{ijk} (\delta \psi_j - \delta \psi_i) ,
\label{eq:2-simplex_rewritten}
\end{align}
\new{by using the symmetry $B_{ijk} = B_{ikj}$ and using $( \delta \psi_j + \delta \psi_k - 2 \delta \psi_i) = ( \delta \psi_j - \delta \psi_i) + ( \delta \psi_k - \delta \psi_i) $. The phase difference in expression~\eqref{eq:2-simplex_rewritten} is similar to the pairwise case, and this allows us to naturally generalize the Laplacian formalism to 2-simplex interactions. }

Indeed, the 2-degree $K^{(2)}_{i}$ of node $i$, i.e. the number of distinct 2-simplices that node $i$ is part of, is $K^{(2)}_{i} = \frac{1}{2!} \sum_{j,k=1}^N  B_{ijk}$ 
where the factor $2!$ ensures each 2-simplex is counted only once. For example, in Fig.~\ref{fig:fig1simplicialnet}(a), the red node has a degree of order 2 $K^{(2)}_{i}=4$. The adjacency matrix  of order 2, whose entries $A^{(2)}_{ij}$ represent the number of 2-simplices shared by the pair $(i,j)$, is $A^{(2)}_{ij} = \sum_{k=1}^N  B_{ijk}$
which is a natural generalization of the usual pairwise adjacency matrix $ A$. Indeed, just as $A_{ij}=1$ if $i$ and $j$ are part of a common 1-simplex interaction but 0 otherwise, $A^{(2)}_{ij}=n$ if $i$ and $j$ are part of $n$ common (but distinct) 2-simplex interactions, but 0 otherwise. 

With these definitions in hand, we can now perform the second important step of our procedure which will take us straight to the Laplacian. We rewrite the 2-simplex interaction term~\eqref{eq:2-simplex_rewritten} as follows
\begin{align}
\delta \dot \psi_i &= \frac{\gamma_2}{\langle K^{(2)} \rangle} \sum_{j,k=1}^N B_{ijk}  (\delta \psi_j - \delta \psi_i) \\
&= \frac{\gamma_2}{\langle K^{(2)} \rangle} \left[ \sum_{j=1}^N A^{(2)}_{ij}  \delta \psi_j -  \delta \psi_i 2! K^{(2)}_{i} \right] , \\
&= \frac{\gamma_2}{\langle K^{(2)} \rangle} \sum_{j=1}^N \left[  A^{(2)}_{ij} - 2 K^{(2)}_{i} \delta_{ij} \right] \delta \psi_j , \\
&= - \frac{\gamma_2}{\langle K^{(2)} \rangle} \sum_{j=1}^N L^{(2)}_{ij} \delta \psi_j , \label{eq:2_laplacian_sys} 
\end{align}
where we obtained the last line by defining the Laplacian of order 2 as in Eq.~\eqref{eq:d-laplacian}, as a natural generalization of the usual pairwise Laplacian. 
With Eq.~\eqref{eq:2_laplacian_sys}, we have explicitly shown the link between the structure of the 2-simplex interactions and the dynamics of the oscillators, by casting it into a Laplacian form. 
Notice that the two essential analytical steps that allowed us to write the Laplacian of order 2 can be straightforwardly generalized at each order. In Appendix~\ref{app:laplacian} we show the same procedure for order 3 and generic order $d$. 

\new{It is important to mention another relevant feature of our formalism. In Eq.~\eqref{eq:system}, we considered the interactions at orders higher than 1 occurring by means of coupling functions designed to be natural generalizations of that at the first order. Specifically, we chose sinusoidal functions which are symmetric with respect to index $i$ and vanish at synchronization. We note here that this two-step derivation above also allows us to treat any other such choice of coupling function, as showed in App.~\ref{app:coupling_functions}. Indeed, by performing the same first step, one notices that terms that do not contain the phase $i$ vanish, and the terms left yield a fraction $c_0/d$ of the Laplacian of order $d$, where $c_0$ is the integer coefficient that multiplies $\theta_i$ in the coupling function. Physically, this means that these other choices, which arise naturally in phase reduction studies beyond the first approximation, see e.g. Refs.~\cite{ashwin2016identical,leon2019phase}, show a slower convergence to synchronization.  }

\subsection{Stability and Lyapunov exponents}

We are finally able to study the stability of our system of oscillators. The synchronized state is stable if the perturbation $\delta \psi_i$ on each node $i$ converges to zero.

Let us first consider pure $d$-simplex interactions. To this end, we need to solve Eq.~\eqref{eq:d-pert} to obtain the temporal evolution of the perturbation. To do so we can make use of the Laplacian eigenvalues $\Lambda_{\alpha}^{(d)}$ and eigenvectors $\phi_{\alpha}^{(d)}$ defined by $\sum_{j=1}^N L^{(1)}_{ij} (\phi_{\alpha}^{(d)})_j = \Lambda_{\alpha}^{(d)}  (\phi_{\alpha}^{(d)})_i$, with $\alpha=1,\dots, N$. Indeed, this eigenbasis can be used to project the perturbation vector $\delta \psi_i (t) = \sum_{\alpha=1}^{N} c_{\alpha} \exp( \lambda_{\alpha}^{(d)} t)  \, \,   \phi_{\alpha}^{(d)} $, where the $c_{\alpha}$ are real constants. By plugging this solution into system~\eqref{eq:d-pert}, we can decouple our system of $N$ equations and obtain the $N$ Lyapunov exponents of the synchronized state
\begin{equation}
\lambda^{(d)}_{\alpha} = - \frac{\gamma_d}{\langle K^{(d)} \rangle }  \Lambda^{(d)}_{\alpha}.
\end{equation}
The Lyapunov exponents are a measure a stability: the system with pure $d$-simplex interactions is stable if all values $\lambda^{(d)}_{\alpha}$ are negative, so that the perturbations $\delta \psi_i$ tend to zero over time. By convention, the Lyapunov exponents are ordered: $\lambda^{(d)}_1 \ge \lambda^{(d)}_2 \ge \dots \ge \lambda^{(d)}_N$. The Laplacian eigenvalues are non-negative by definition so that if all $\gamma_d>0$ (attractive coupling), the synchronized state is always stable up to a global phase shift: \mbox{$0=\lambda^{(d)}_1 > \lambda^{(d)}_2$}.

In the multiorder system~\eqref{eq:systemLapl} the stability of the synchronized equilibrium is determined by the interplay of all different orders, as encoded in the multiorder Laplacian. We can then analogously use the spectrum and eigenbasis of $\lap^{\text{(mul)}}$, hence obtaining the $N$ Lyapunov exponents
\begin{equation}
\lambda^{\text{(mul)}}_{\alpha} = - \Lambda^{\text{(mul)}}_{\alpha}.
\end{equation}
The Laplacian eigenvalues are non-negative by definition so that if all $\gamma_d>0$ (attractive coupling), the synchronized state is always stable. The Lyapunov exponent that determines the long-term behavior of the second Lyapunov exponent, $\lambda_2^{\text{(mul)}}$, i.e. the smallest non-zero one.  Its value determines the resilience of the system to perturbations, i.e. how fast the system comes back to the stable state after a perturbation. In particular, the more negative the $\lambda_2^{\text{(mul)}}$ is, the more stable is the synchronized state.

\section{Stability in all-to-all higher-order networks} \label{sec:all-to-all}


In this section we analyze a simple case -- dubbed \textit{higher-order all-to-all} -- which can be solved analytically. This setting is a generalization of the usual ``all-to-all'' (or global) coupling scheme in traditional networks. Indeed, in networks with pairwise interactions, \textit{all-to-all} coupling indicates that every possible pairwise interaction takes place. Similarly, in a network with higher-order interactions, \textit{higher-order all-to-all} indicates that every possible $(d+1)$-oscillator interaction occurs, for all orders $d$. 

\new{This setting is the only one that has been studied analytically, with a main focus on cluster states. In this homogeneous case, each term of order $d$ in the original system~\eqref{eq:system} can be written in terms of the order parameter amplitude $R_1$ and its phase $\Phi_1$ as, up to a normalization factor
\begin{equation}
\dot \theta_i = \gamma_d \, R_1^d \, \sin [d(\Phi_1 - \theta_i)] = \gamma_d \ima[Z_1^d e^{-i\theta_i}]
\label{eq:nonlinear_meanfield}
\end{equation}
which makes apparent the driving by the meanfield that is now nonlinear. Effectively, each oscillator is driven by the same meanfield with strength $R_1^d$ and with a $d$-th harmonic. Pure harmonics are known to yield stable cluster states, which can be checked via a self-consistency argument. Take $d=2$, then Eq.~\eqref{eq:nonlinear_meanfield} has two stable fixed points with distance $\pi$, and that 2-cluster state has $R_1>0$ which keeps driving the system. 
}

\new{These considerations were used in \cite{skardal2019abrupt} for the pure triplet case,  to study cluster states and abrupt transition in the thermodynamic limit. Still in the pure triplet case, \cite{komarov2015finitesizeinduced} shows that although the incoherent state is stable in the thermodynamic limit, finite-size effects can destabilize it and yield cluster states. In~\cite{skardal2019higher}, the authors, combining interactions of orders up to 4, unveil the emergence of an abrupt transition to synchronization in the thermodynamic limit. Finally, in~\cite{gong2019lowdimensional}, the authors extended the Watanabe-Strogatz low-dimensional description to any pure higher harmonics $l$ for the general system $\dot \theta_i = \omega + \ima[H e^{-i l \theta_i}]$, where $H$ is a function of the generalized order parameters, which can take the form~\eqref{eq:nonlinear_meanfield} and yield higher-order interactions of any order. With their framework, the authors tracked the basins of attraction of the clusters in cluster-states. While this treatment does not primarily focus on higher-order interactions, it can be related via the nonlinear meanfield coupling.  }

\new{Here, in contrast with these studies, we focus on the stability of full synchrony and provide a full analytical description of its spectrum. In addition, we investigate the effect considering a variable number of orders, mixing attractive and repulsive couplings, and decaying coupling strengths as observed in phase reduction studies. In particular, we show that, due to the absence of complex topology, the Lyapunov spectrum of the full system reduces to a linear combination of those at each pure order.
We remark that this is not true in complex topologies, for which the aggregated spectrum is determined by a nonlinear combination of each order.}

\begin{figure*}[htb]
	\centering
	\includegraphics[width=0.99\linewidth]{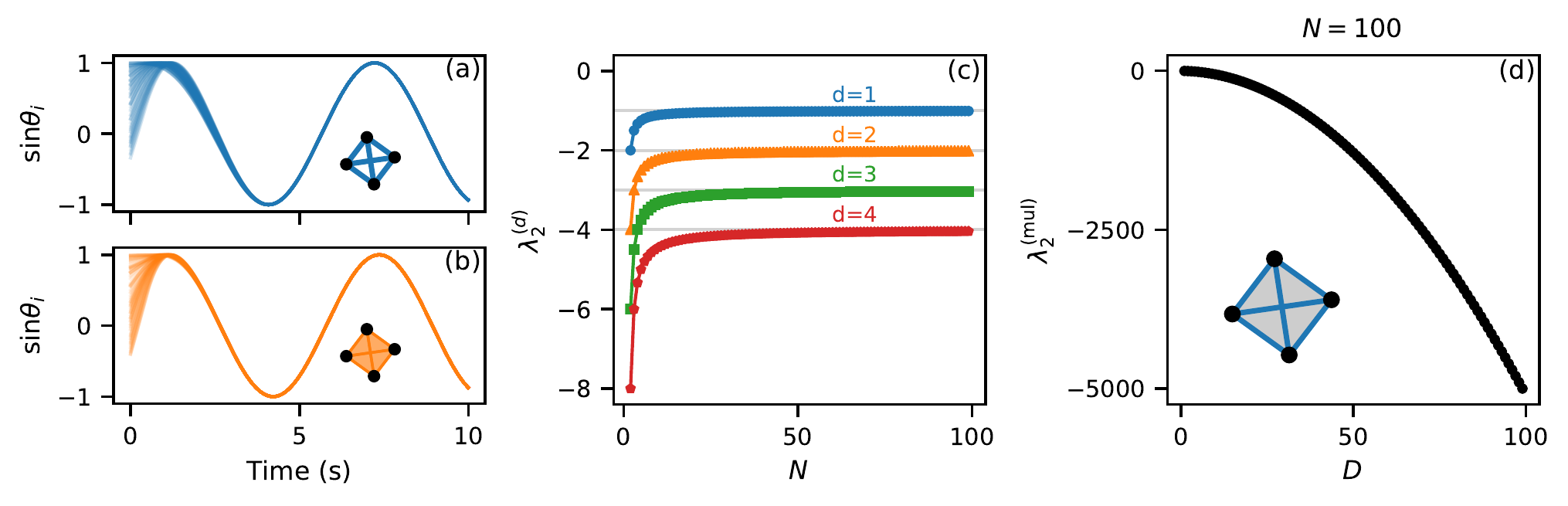}
	\caption{\textbf{Higher-order all-to-all: attractive coupling.} Higher-order interactions increase the stability of synchronization. Synchronisation of $N=100$ oscillators over time with (a) only 1-simplex (2-oscillator) interactions and (b) only 2-simplex (3-oscillator) interactions, obtained by numerical integration. The oscillators with only 2-simplex interactions synchronize faster than those with only 1-simplex ones. This is confirmed by (c): the analytical second Lyapunov exponent $\lambda_2^{(d)}$ of Eq.~\eqref{eq:lyap_dmax_ata_attract} as a function of network size $N$, at each order $d=1,\dots,4$. The Lyapunov exponent is more negative for larger orders $d$ of interaction. (d) Lyapunov exponent of the full system, as a function of $D$, the largest order taken into account. The higher the order of interactions taken into account, the more negative the Lyapunov exponent is, i.e. the most stable the synchronized state is. Parameters are $N=100$ and  $\gamma_d=1$ for all $d$.}
	\label{fig:all-to-all-lyap_partial}
\end{figure*}

\subsection{Higher-order Laplacians are proportional to the traditional pairwise Laplacian}
\label{subsec:all-to-all-Lap}

We now show that at each order $d$, the higher-order all-to-all Laplacian  $\lap^{(d)}$ is proportional to the usual pairwise all-to-all Laplacian $\lap^{(1)}$, defined by $L_{ij}^{(1)} = K_{ij}^{(1)} \delta_{ij} - A_{ij}^{(1)} $. Consequently, since the analytical spectrum of the latter is known, the analytical spectrum of $\lap^{(d)}$ can also be obtained, and in turn, that of $\lap^{\text{(mul)}}$. 

\new{First, we need to explicitly write down our degree and adjacency matrices defined in general in Eqs.~\eqref{eq:d-degree} and \eqref{eq:d-adj}. In the all-to-all case, the degree reduces to $K_i^{(d)} = {N-1 \choose d}$, in combinatorics notation. This, for order 1, yields the usual $K_i^{(1)} = N-1$. The adjacency matrix reduces to $A_{ij}^{(d)}= {N-2 \choose d-1} (1-\delta_{ij})$, which yields the usual  $A_{ij}^{(1)} = 1-\delta_{ij}$ at order 1.
Additionally, the following identities will prove useful in the next section, to relate the quantities at order $d$ to their traditional counterparts (at order 1)
\begin{align}
A_{ij}^{(d)} &= [(N-2) \cdots(N-d) / (d-1)!] \, A_{ij}^{(1)} \label{eq:adj_d_1} ,\\
K_{i}^{(d)} &= [(N-2) \cdots(N-d) / d!] \,  K_{i}^{(1)}  \label{eq:deg_d_1} .
\end{align}
Now that we have these expressions, we have all we need to write an explicit formula for the multiorder Laplacian~\eqref{eq:d-laplacian} in the next section. In the higher-order all-to-all setting, all nodes have the same degree of order $d$. Hence, in this section, we write $K_{i}^{(d)}$ and $K^{(d)}$ interchangeably. More details are provided in Appendix~\ref{app:all-to-all}.}

\textbf{Pure $d$-simplex interactions.} For the general case of order $d$, analogously injecting expressions~\eqref{eq:adj_d_1} and~\eqref{eq:deg_d_1} into definition~\eqref{eq:d-laplacian} yields
\begin{align}
L_{ij}^{(d)} &= d K_{i}^{(d)} \delta_{ij} - A_{ij}^{(d)} , \\
&= [(N-2) \cdots(N-d) / (d-1)!] \, \, [K_{i}^{(1)} \delta_{ij} - A_{ij}^{(1)}] , \\
&= [(N-2) \cdots(N-d) / (d-1)!] \, \, L_{ij}^{(1)} ,
\end{align}
which can be rewritten
\begin{equation}
L_{ij}^{(d)} =  \frac{d \,  K^{(d)}}{N-1} \, \, L_{ij}^{(1)}.
\label{eq:laplacian_d_1}
\end{equation}
Hence, we have shown that the Laplacian of order $d$ is proportional to the usual pairwise Laplacian, for any order $d$. In addition, Eq.~\eqref{eq:laplacian_d_1} indicates that $L_{ij}^{(d)}$ is linearly growing with the order of the interactions $d$, and with $K^{(d)}$, which the number of $d$-simplex interactions that each oscillator has. 

We will discuss the implications that these dependencies have on the stability in more detail in Sec.~\ref{sec:spectrum}. For now, we only stress that the analytical formula~\eqref{eq:laplacian_d_1} serves as a limit case to understand the behavior of $L_{ij}^{(d)}$ in more complex coupling schemes than the higher-order all-to-all scheme. Examples of those will be investigated in Sec.~\ref{sec:complex}, in which a full analytical derivation is not always possible.

\textbf{multiorder interactions.} Oscillators in the higher-order network have multi-oscillator interactions with the other oscillators at all orders $d=1, \dots, D$. In general, the stability of the synchronized solution is determined by the eigenvalues of the multiorder Laplacian $\lap^{\text{(mul)}}$, in Eq.~\eqref{eq:laplacian_tot},  as we have shown in Sec.~\ref{sec:laplacian}. In the higher-order all-to-all setting, by injecting Eq.~\eqref{eq:laplacian_d_1}, this Laplacian reduces to
\begin{equation}
L_{ij}^{\text{(mul)}} =  \left( \sum_{d=1}^{\new{D}} \frac{\gamma_d \, d}{N-1} \right) \, \, L_{ij}^{(1)} .
\label{eq:laplacian_tot_ata}
\end{equation}

Hence, the stability of the synchronized solution is only determined by the usual pairwise Laplacian, as well as by the strength $\gamma_d$ of each interaction of order $d$, and by the order $d$ of those interactions.  We give the full analytical spectrum of $\lap^{\text{(mul)}} $ for the higher-order all-to-all case in the next section. \new{We will see that, as a consequence of the proportionality between $L_{ij}^{(d)}$ and $L_{ij}^{(1)}$, the multiorder eigenvalue spectrum is just a linear combination of the spectra at each order. However, this is only true in the all-to-all scheme which overlooks the complexity of real-world topologies.}

\begin{figure*}[htb]
	\centering
	\includegraphics[width=0.99\linewidth]{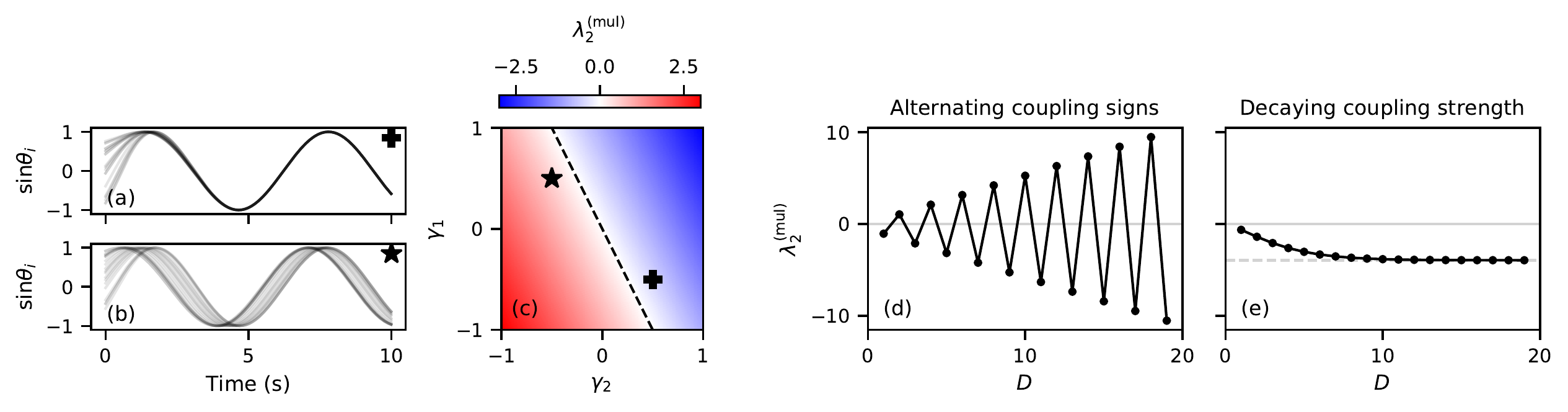}
	\caption{\textbf{Higher-order all-to-all: (a)-(d) interplay of attractive and repulsive coupling orders, and (e) decaying coupling strength.} Phases over time with (a) which synchronize even with repulsive 1-simplex interactions due to attractive 2-simplex interactions ($\gamma_1=-0.5$, $\gamma_2+-0.5$, black plus in (c)), but do not synchronize with attractive 1-simplex interactions because of repulsive 2-simplex interactions ($\gamma_1=+0.5$, $\gamma_2=-0.5$, black star in (c)). Times series are numerically integrated on a simplicial complex of $N=100$ nodes. (c) Analytical, non-zero, multiorder Lyapunov exponent (with $D=2$) for a range of 1- and 2-simplex coupling strengths $(\gamma_1, \gamma_2)$. Positive and negative coupling strengths correspond to attractive and repulsive coupling, respectively. Negative values (blue) and positive values (red) of $\lambda_2^{\text{(mul)}}$ indicate instability and stability of the synchronized state, and confirms numerics in (a) and (b). The Lyapunov exponent vanishes when $\gamma_1 + 2 \gamma_2 = 0$ (dashed black). (d) Analytical, non-zero, multiorder Lyapunov exponent as a function of the largest order taken into account $D$, for attractive and repulsive coupling at even and odd orders, respectively. The synchronized state changes its stability as higher-orders are taken into account. (c) Analytical, non-zero, multiorder Lyapunov exponent as a function of  $D$, for decaying coupling strengths $\gamma_d = \gamma_1^d$, with $ \gamma_1=0.6$. The Lyapunov exponent converges to a negative value as higher-order interactions are taken into account.}
	\label{fig:fig3all-to-all-lyapalternating}
\end{figure*}

\subsection{Spectrum and stability} \label{sec:spectrum}

In this section, we give an analytical formula for the eigenvalues of the higher-order all-to-all Laplacian matrix, and consequently for the Lyapunov exponents which determine the stability of the synchronized solution. We remind the reader that the attractiveness or repulsiveness of interaction at a given order $d$ is determined by the sign of the coupling strength $\gamma_d$: positive and negative coupling strengths correspond to attractive and repulsive interactions, respectively. We consider different scenarios, where we tune at will the sign and intensity of the coupling strengths $\gamma_d$.

\textbf{Attracting couplings at all orders.} We have shown in the previous section that the $d$-order Laplacians are intrinsically connected to the usual pairwise Laplacian $\lap^{(1)}$. Therefore, this latter shapes the Laplacian spectrum at each order $d$, and consequently also at the multiorder.  For higher-order all-to-all networks, the spectrum of $\lap^{(1)}$ is degenerate and given by 
\begin{equation}
\Lambda^{(1)}_{1} = 0 \qquad \Lambda^{(1)}_{2,\ldots,N} = N
\end{equation}
from which we derive the Lyapunov exponents of each order $d$:
\begin{equation}
\lambda^{(d)}_{1} = 0 \qquad \lambda^{(d)}_{2,\ldots,N} = - \gamma_d \, d \, \frac{ N }{N-1}.
\end{equation}
The second Lyapunov exponent is reported in Fig.~\ref{fig:all-to-all-lyap_partial}(c) for different values of $d$ as a function of network size $N$. It appears clear that interactions of higher orders $d$ stabilize the synchronized state more (more negative $\lambda^{(d)}_{2}$ as $d$ increases). Interestingly, this is true despite each order being given an equal weight through the normalization in system~\eqref{eq:system}.  This is illustrated with numerical simulations for the pure orders $d=1$ and $d=2$, respectively shown in Figs.~\ref{fig:all-to-all-lyap_partial}(a) and~\ref{fig:all-to-all-lyap_partial}(b). Indeed, trajectories converge faster with pure 2-simplex interactions than with pure pairwise ones. 

Then, combining all orders, we obtain the multiorder Lyapunov exponents 
\begin{equation}
\lambda^{\text{(mul)}}_{1} = 0 \qquad \lambda^{\text{(mul)}}_{2,\ldots,N} = - \frac{N}{N-1} \sum_{d=1}^{\new{D}} \gamma_d \, d .
\label{eq:lyap_dmax_ata_attract}
\end{equation}
From this formula, we see that the more orders are taken into account, i.e., the more $D$ is increased, the more negative $\lambda^{(\text{mul})}_{2}$ is, as shown in Fig.~\ref{fig:all-to-all-lyap_partial}(d). Physically, additional attractive higher-order interactions tend to stabilize the synchronized state, \new{as expected}. 

\begin{figure*}[htb]
	\centering
	\includegraphics[width=0.8\linewidth]{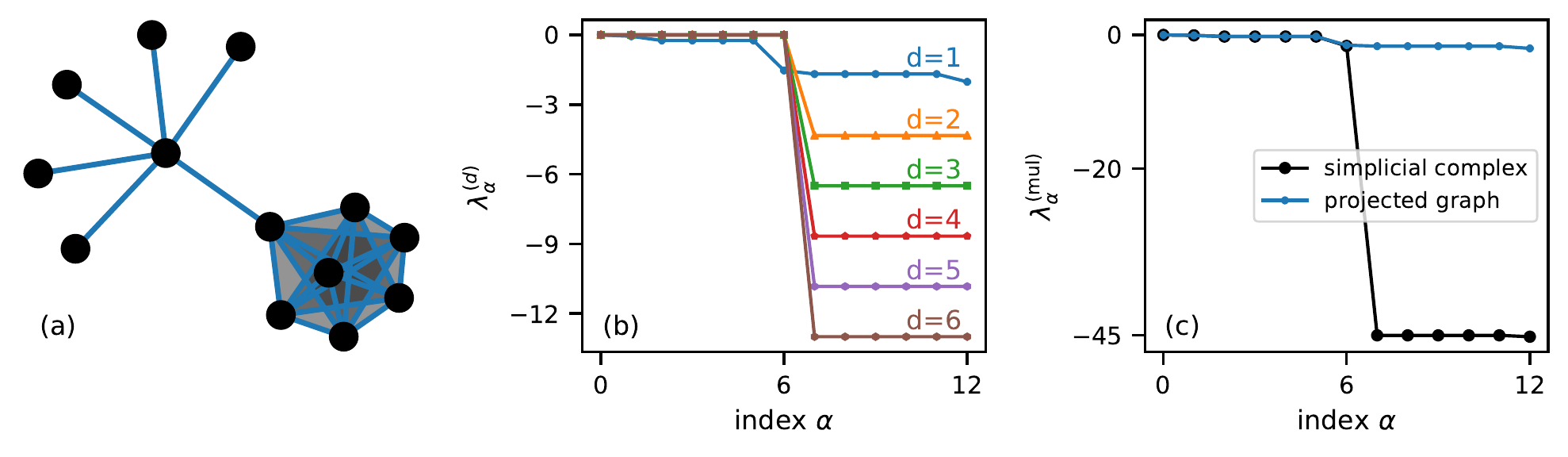}
	
	\caption{\textbf{Synchronisation in the star-clique simplicial complex (a) with $N_s=6$ and $N_c=7$.} The highest-order of interaction is $D=6$. In the clique, faces are part of simplices of 5 different orders, and hence they are depicted in gray. (b) Lyapunov spectra for each case of pure $d$-simplex interactions. (c) Lyapunov spectrum with higher-order interactions combined (black), $\gamma_d=1$ for all $d$, compared to the traditional pure 1-simplex case (blue).}
	\label{fig:starcliquem7k6spectrumg11g21}
\end{figure*}

\textbf{Attractive even orders and repulsive odd orders.} Here, we apply our analytical framework to investigate the interplay between attractive and repulsive interactions. For simplicity, we study the case where attractive and repulsive relationships are associated with different interaction orders. \new{Mixing attractive and repulsive couplings has been previously investigated, motivated by, e.g., analogies with inhibitory and excitatory connections between neurons~\cite{hong2011}.}

First, we restrict ourselves to only 1-simplex (2-oscillator) and repulsive 2-simplex (3-oscillator) interactions. Depending on the sign of $\gamma_1$, 1-simplex interactions are attractive ($\gamma_1>0$) or repulsive  ($\gamma_1<0$). The attractiveness or repulsiveness of 2-simplex interactions depends identically on the sign of $\gamma_2$. Physically, attractive interactions favor synchronization by stabilizing the synchronized state. By contrast, repulsive interactions will favor incoherence by destabilizing the synchronized state.  The result of this interplay between the interactions at various orders depends on the respective coupling strengths $\gamma_d$:  $\lambda^{(\text{mul})}_{2} = - \frac{N}{N-1} (\gamma_1 + 2 \gamma_2)$ which is negative if $\gamma_1 + 2 \gamma_2 > 0$. The value of $\lambda^{(\text{mul})}_{2}$ is shown in Fig.~\ref{fig:fig3all-to-all-lyapalternating}(c) for a range of positive and negative values of $\gamma_1$ and $\gamma_2$. We see that synchronization can be stable even if the traditional pairwise (1-simplex) interactions are repulsive, as long as 3-oscillator (2-simplex) interactions are attractive enough to counterbalance them, as illustrated in  Fig.~\ref{fig:fig3all-to-all-lyapalternating}(a). This highlights a potential benefit of considering higher-order interactions: indeed they might stabilize the systems, in cases when the purely pairwise system is unstable. This confirms a similar result obtained in~\cite{skardal2019abrupt} for phases oscillators with distributed frequencies. In addition, attractive pairwise interactions might be outplayed by repulsive 3-oscillators ones. This can result in an unstable synchronized state, as illustrated in~Fig.~\ref{fig:fig3all-to-all-lyapalternating}(b).

Second, we generalize to all orders up to $D$ with alternating signs. Specifically, we consider all interactions of an even and odd order $d$ to be attractive and repulsive, respectively, by setting
\begin{equation}
\gamma_{2n} = -1 \qquad \gamma_{2n+1} = +1 .
\end{equation}
Now, the Lyapunov exponent is given by the alternating series
\begin{equation}
\lambda^{(\text{mul})}_{2} = - \frac{N}{N-1} \sum_{d=1}^{D} d \, (-1)^{d+1} ,
\end{equation}
which diverges as $D \to \infty$ (which requires  $N \to \infty$). For increasing but finite values of $D $, however, the Lyapunov exponent alternates between positive and negative values, as illustrated in Fig.~\ref{fig:fig3all-to-all-lyapalternating}(d). In other words, if the highest order of interaction considered is odd, the synchronized state is stable. However, if it is even, the synchronized state is unstable. This is due to the fact that the contribution to $\lambda^{\text{(mul)}}_{2}$ of each order $d$ is proportional to $d$: adding one repulsive or attractive order outplays all lower-order interactions. Finally, we note that the factor $N/(N - 1) \to 1$ in the limit of large networks $N \to \infty$, so that it can be seen as a finite size correction factor.

\textbf{Weaker higher orders: link to phase reduction.} Here, we make a brief link between our formalism and the higher-order phase reductions approaches developed for example in~\cite{bick2016chaos,bick2016chaotic,leon2019phase,rosenblum2019numerical}. In these phase reductions studies, the authors obtain a phase model from an original network of nonlinear oscillators, by performing a sophisticated perturbative expansion in a small parameter. This small parameter is usually linked to the original pairwise coupling strength $\gamma_1$. The authors find that, at higher orders in the expansion, i.e. at higher powers of $\gamma_1$, there appear terms including higher-order interactions, i.e. interactions between more than 2 phases. 

Motivated by these studies, we consider a scenario of decaying coupling strengths. Specifically, we set $0 \le \gamma_1<1$, and couplings at higher orders as powers of the pairwise coupling strength
\begin{equation}
\gamma_d = \gamma_1^d ,
\end{equation}
which means that interactions between many oscillators are weak. In this case, the stability is given by the series
\begin{equation}
\lambda^{(\text{mul})}_{2} = - \frac{N}{N-1} \sum_{d=1}^{D} d \, \gamma_1^d <0.
\end{equation}
which is always negative, and hence the stability of the synchronized state is ensured. In addition, this is a geometric series that converges to $-\gamma_1 / (\gamma_1 - 1)^2$ when $D$ (and hence $N$) tends to infinity. This convergence is illustrated in Fig.~\ref{fig:fig3all-to-all-lyapalternating}(e). Even though the qualitative result, i.e. stability, was expected since interactions at all orders are attractive, such a quantitative result has more predictive power. In particular, we note that $-\gamma_1 / (\gamma_1 - 1)^2 \to -\infty$ as we go away from the domain of validity of the perturbative regime $\gamma_1 \to 1$.

\section{Stability in higher-order networks with arbitrary topology} \label{sec:complex}

In this section, we apply our multiorder Laplacian framework to simplicial complexes with complex heterogeneous structures. We consider two cases: first, a toy model --the star-clique model-- and second, a real network, i.e. a macaque brain dataset. 

\begin{figure*}[htb]
	\centering
	\includegraphics[width=0.9\linewidth]{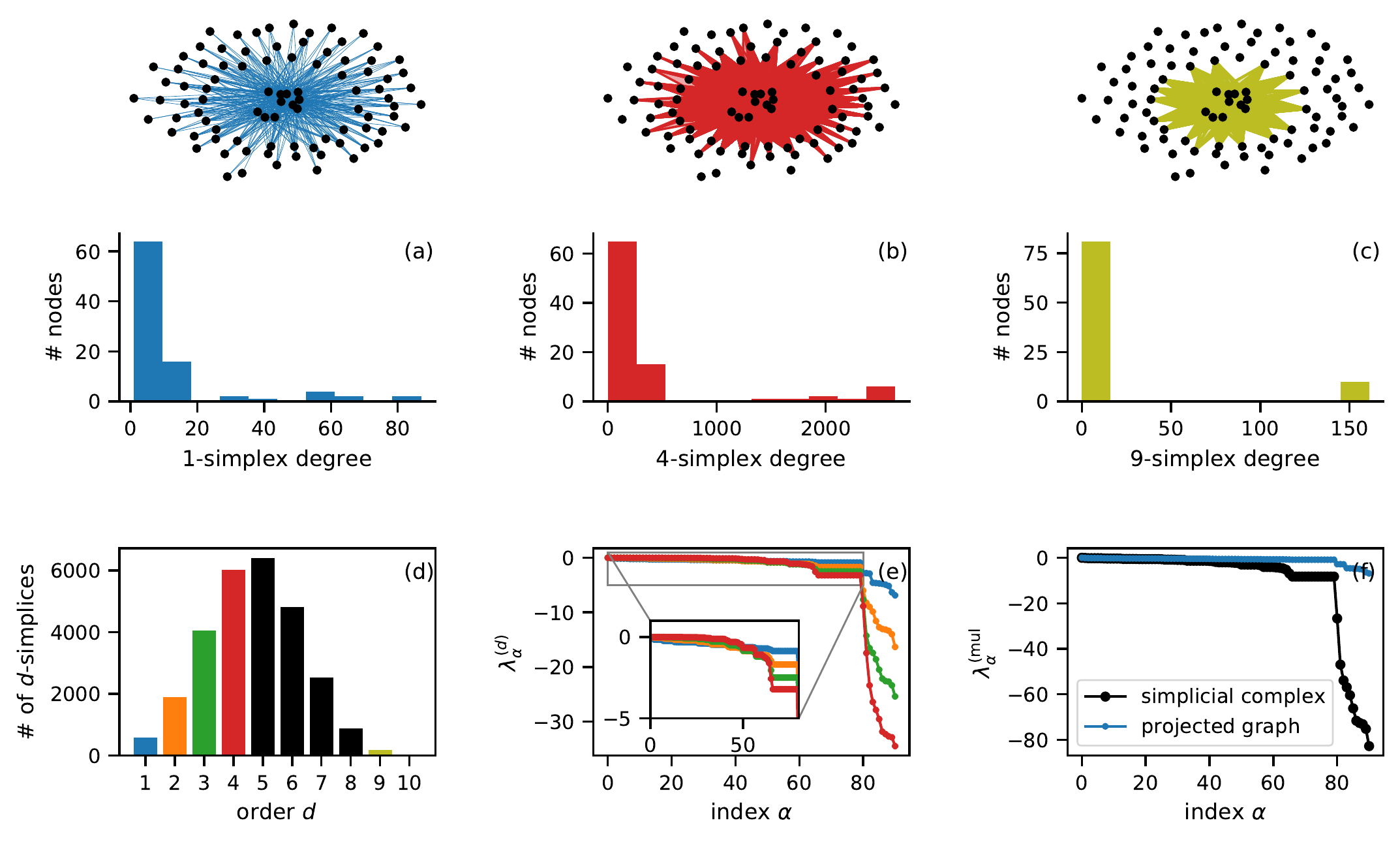}
	\caption{\textbf{Real macaque brain dataset: structure and dynamics.} Panels (a), (b) and (c) depict the simplicial complex as it appears at orders $d=1$, 4 and 9 respectively, and they report the associtated distributions of degree $K_i^{(d)}$ across nodes. At each order, most nodes have zero or few interactions, but a few nodes have many. (d) Number of $d$-simplices in the simplicial complex, for each $d$. The highest order of interaction in the network is $D=10$. (e) Lyapunov spectra with only 1-simplex interactions, and with higher-order interactions (with $d$ up to $D=4$). (f) Comparison of the Lyapunov spectra:  multiorder (black), and projected graph (blue), i.e. $D=1$. The spectrum is affected by higher-order interactions.}
	\label{fig:macaque}
\end{figure*}

\subsection{The star-clique model}

As a first example of complex topology with higher-order interactions, we consider a toy model that we call {\it star-clique}~\cite{carletti2020random}. We make use of this model in order to observe in a simple case the multiorder Laplacian properties and thus the stability conditions for synchronization. The star-clique, as its name indicates, is composed of two subnetworks: a star of size $N_s$ and a clique of size $N_c$, which we treat analogously to a higher-order all-to-all network of size $N_c$ from the previous section. The subnetworks are connected only by one link from the center of the star to a single node of the clique, as illustrated in Fig.~\ref{fig:starcliquem7k6spectrumg11g21}(a). By construction of this simplicial complex, the two subnetworks are almost disconnected, and work almost as two independent systems. We will see that, however, that at each pure order $d\ge2$ the subnetworks are actually disconnected. This allows us to use our analytical understanding from the previous section to comprehend the present case with complex topology.

We start by analyzing the structure of the 1-simplex interactions. The Lyapunov spectrum of the pure 1-simplex interactions is shown in blue in Fig.~\ref{fig:starcliquem7k6spectrumg11g21}(b). The spectrum reflects the quasi-independence of the two subnetworks. Indeed, since the star and the clique are almost independent, the adjacency matrix $A_{ij}$ is a slightly perturbed matrix with two independent blocks, one for each subnetwork. The same holds for the Laplacian matrix, and hence its eigenvalues and the Lyapunov exponents. We know that the $N_s$ Laplacian eigenvalues  $\Lambda_{\alpha}^{(1)}$ of a star network with pairwise interactions are given by $\{0, 1, \dots, 1, N_s+1\}$. We also know that the $N_c$ Laplacian eigenvalues $\Lambda_{\alpha}^{(1)}$ of a pairwise all-to-all network are given by  $\{ 0, N_c, \dots, N_c\}$. Hence, the spectrum of $N$ eigenvalues $\Lambda_{\alpha}^{(1)}$ of the 1-simplex Laplacian $\lap^{(1)}$ is the union of both spectra, only slightly perturbed. This can be seen on Fig.~\ref{fig:starcliquem7k6spectrumg11g21}(b) for the Lyapunov exponents, i.e. scaled Laplacian eigenvalues (blue symbols). 

Moving to higher-order interactions, $d\ge2$, the star and the clique actually are disconnected. In fact, in these pure $d$-simplex cases, there is no star, since it has no higher-order interactions. Hence the higher-order Laplacians only ``see'' the clique. This implies that the spectrum of $L_{ij}^{(2)}$ is the union of $N_s$ zeros and the spectrum of a higher-order all-to-all that we derived analytically in the previous section: one element zero and the others with value $- \gamma_2 2  N_c / (N_c-1) $. Analogously at orders $d$ higher than 2, the spectrum is given by $N_s + 1$ zeros and $N_c -1$ elements $- \gamma_d d  N_c / (N_c-1) $. See Fig.~\ref{fig:starcliquem7k6spectrumg11g21}(b) for the spectrum at different orders $d$.

For what concerns the spectrum of the total Laplacian, it is not exactly the sum of the spectrum of order $d$, as in the all-to-all case, but almost. Indeed, the only thing that distinguishes it from the previous case, is that the pairwise Laplacian of the star-clique network is not composed of two disconnected blocks, but almost. Let us observe that, if the pure orders $d \ge 2$ can be simply described by the all-to-all higher order spectrum, the order $d=1$ (traditional pairwise) and the multiorder, reflect the structure of the two almost disconnected subgraphs, namely the star and the clique. Their spectrum is compared in Fig.~\ref{fig:starcliquem7k6spectrumg11g21}(c). It is important to notice that $\lambda_2^{(1)}$ and $\lambda_2^{(\text{mul})}$ are identical. The two spectra are however globally different and in particular $\lambda_N^{(1)}$ and $\lambda_N^{(\text{mul})}$ are very have very different values, suggesting that in the repulsive case (all $\gamma_d<0$) the two Laplacians would yield dynamics with very different timescales.

\subsection{Real network: Macaque brain} \label{sec:macaque}

In this section, we demonstrate the use of our multiorder Laplacian framework on a real dataset. We use a macaque brain dataset publicly available at~\cite{braindata}, consisting of 91 nodes and 628 edges. \neww{The nodes represent cortical areas and the edges links among those, experimentally determined by retrograde tracing~\cite{markov2014anatomy}.} We assume that each fully connected ($d+1)$-node clique captures multiorder interactions that can be described as a $d$-simplex, and we study the dynamics of \eqref{eq:system} on top of the resulting simplicial complex. The obtained simplicial complex as it appears at orders 1, 4 and 9 is depicted in Fig.~\ref{fig:macaque}, respectively (a), (b) and (c). \new{A whole spectrum of models of coupled oscillators has been used to study the synchronization of neurons in the brain. \neww{Here we do not intend this to be a realistic model of a functional brain, but we rather use this dataset as empirical ground to test our theoretical toy model of synchronization in systems with higher-order interactions}. While there are clear limitations in applying extremely simplified neuronal models to capture real brain dynamics, idealized phase models are still interesting, as they can sometimes capture well more realistic dynamics like integrate-and-fire ones~\cite{politi2015equivalence}}.

The full distribution of the number of $d$-simplices present in the simplicial complex is shown in Fig.~\ref{fig:macaque}(a). The highest order of interaction is 10. As expected, the distribution follows a bell curve: the simplicial complex contains only less than 1000 1-simplices, i.e. 2-oscillator interactions, and 10-simplices, i.e. 11-oscillator interactions, but has over 6000 5-simplices. In addition to the distribution of $d$-simplices across orders $d$, the dynamics is affected by the structure of the $d$-simplex interactions at each order $d$. Specifically, we wonder how 1-, 2-, \dots, and $d$-simplices are distributed among nodes. It turns out that at each order $d$, the interactions are very centralized: only few nodes have many $d$-simplex interactions whereas the majority of the nodes have very few or no $d$-simplex interaction at all. This is shown for order 4 and 9 in Figs.~\ref{fig:macaque}(b) and \ref{fig:macaque}(c), respectively. The higher the order $d$ is, the more nodes have zero $d$-simplex interactions. 

How is the stability of the synchronized state affected by this inter- and intra-order structure? To assess this, we compute the Laplacian~\eqref{eq:d-laplacian} at each order $d$ as well as the multiorder Laplacian~\eqref{eq:laplacian_tot}, and obtain the Lyapunov exponents from the eigenvalue spectra. First, we assess the stability of each pure case, at orders $d$ from 1 to 4, by computing their respective Lyapunov spectra shown in Fig.~\ref{fig:macaque}(e). Orders highers than 4 are not shown for visualization purposes. These spectra reflect the structure we  described above: the higher the order $d$, the more nodes that have no $d$-simplex interactions, i.e. that are disconnected of the main component in the pure $d$-simplex network. These disconnected nodes translate into eigenvalues of value 0 in the spectra. 
We observe also two other stages of the spectra: a first step where higher orders correspond to lower Lyapunov exponents, which is similar to what happens for the star-clique model because of the clique subnetwork and may reflect, also in this example, the aggregated nature of the network at each order $d$. Then, we observe a strong descending behavior for the last eigenvalues, which confirms the stronger stability of the higher orders.

Finally, we compare the stability of the full system with combined higher-order interactions against what is obtained on the corresponding projecting graph, where nodes coupled at any order are linked with a pairwise edge. This is illustrated in Fig.~\ref{fig:macaque}(f).

\section{Summary and conclusions} \label{sec:conclusion}

In summary, in this work we have studied the effects of higher-order interactions, i.e. multi-oscillator interactions, on the synchronization of identical phase oscillators on hypergraphs. We considered interactions up to any order in an extension of the Kuramoto model by introducing a multiorder Laplacian, which makes the system amenable to analytical treatment, and determines the stability of the fully synchronized state. In particular, we obtained a quantitative measure of that stability by computing the Lyapunov exponents of the state, that are proportional to the Laplacian spectrum. We showed applications of the multiorder Laplacian in settings of increasing complexity. \new{We emphasize that such analytical treatment is especially important for higher-order systems, as numerical simulations become slower with each additional order of interactions, quickly becoming unfeasible.}

As a first example, we considered a case that is fully tractable analytically: the higher-order all-to-all setting. In this setting, all nodes interact in all possible $d$-simplex interactions for any order $d$. In this case, we showed that the Laplacian of each order $d$, $\lap^{(d)}$, is proportional to the traditional pairwise Laplacian $\lap^{(1)}$, and to the value of $d$ itself, see Eq.~\eqref{eq:laplacian_d_1}. As a consequence, the stability of the multiorder system can be linearly decomposed into the stability of each of the pure $d$-simplex systems. Indeed, in this special setting the multiorder Lyapunov exponents are merely a weighted sum of the Lyapunov exponents of order $d$. We confirm our findings with numerical simulations. Finally, this fully tractable higher-order all-to-all setting also serves as a limit case to help us understand what happens in more complex topologies. 

In this context, we investigated analytically two additional scenarios. First, we investigated the interplay between orders with attractive interactions ($\gamma_d>0$) and orders with repulsive interactions ($\gamma_d<0$). For example, we showed that their opposite effect on synchronization means that repulsive pairwise couplings can be countered effectively by higher-order attractive couplings. In general, we showed that when odd orders are attractive and even orders are repulsive, taking more higher-orders into account can stabilize or destabilize the system, depending on the highest order considered. Second, to link our work to higher-order phase reduction techniques, we considered attractive coupling strength $\gamma_d$ that decays as the order $d$ increases. We derived the convergence of the Lyapunov exponents as higher orders are considered.

Then, we considered two cases with more complex topologies. First, we applied the multiorder Laplacian framework to a toy model: the star-clique network. With this model, we illustrated spectral properties of the multiorder Laplacian, and of the Laplacian at each order. In that specific topology, the star subnetwork does not contain higher-order interactions and hence does not influence the value of the first nonzero Lyapunov exponent. Higher-order interactions in the clique subnetwork, however, do change the other values of the spectrum, as we showed. Second, we considered a real-world topology: a macaque brain network. In this setting, we show how the complex shape of the multiorder Laplacian spectrum can be understood from the structure of the simplicial complex. More importantly, our analysis confirms changes in all Lyapunov exponents due to the inclusion of higher-order interactions, as compared to only pairwise interactions. Specifically, higher-order interactions tend to stabilize synchronization, with more drastic change on the more negative part of the spectrum, due to the hub-like structure of the dataset. \new{We showed in Appendix~\ref{app:coupling_functions} that, for higher-order complex topologies,  the choice of the coupling function affects the timescales, causing them to reach full synchrony. Yet, the precise effect of different coupling functions~\cite{stankovski2015coupling,stankovski2017} on other dynamical regimes on such systems is an open question.} 

In conclusion, in this paper, we introduced a multiorder Laplacian to assess the stability of synchronization in populations of oscillators with higher-order interactions. Our framework has two main strengths: (i) it is a natural generalization of the traditional and well known pairwise Laplacian framework and (ii) it can be applied to \new{arbitrary hypergraphs describing any structured group interactions up to any order $d$. This is in contrast with previous studies, where analytical insights were provided for the higher-order all-to-all coupling scheme only. Besides, other than its inherent ability to deal with complex topologies, the Laplacian formalism is valid for any $N$, allowing one to investigate finite-size effects.} 

\new{Our framework promises to find wide applicability. Indeed, it has very recently been used to extend the ideas behind the master stability function~\cite{pecora1998master} to simplicial complexes~\cite{gambuzza2020master}. 
Taken together, the multiorder Laplacian is a powerful tool that could be used widely not only to characterize populations of oscillators with higher-order interactions, but also for a wider general analytical treatment of dynamical processes beyond pairwise interactions.}

\subsection*{Acknowledgments} 

The authors would like to thank Duccio Fanelli, Iacopo Iacopini, and Arkady Pikovsky for fruitful discussions, as well as Alain Barrat for feedback on the manuscript. M.L. and G.C. acknowledge partial support from the ``European Cooperation in Science \& Technology'' (COST): Action CA15109. F.B. acknowledges partial support from the ERC Synergy Grant 810115 (DYNASNET).

\appendix

\section{Derivation of the Laplacian formulation}
\label{app:laplacian}

\subsection{Order 3}

We briefly show how to rewrite the 3-simplex interactions of Eq.~\eqref{eq:linearised_system} in a similar way to the 2-simplex case from Sec.~\ref{subsec:laplacian}. Although it is similar, the case of 3-simplices better prepares us for the next step: the general $d$-simplex case.  The first important step is to reduce the expression with 4 phases into an expression of only 2 phases
%
	\begin{align}
	\delta \dot \psi_i =& \frac{\gamma_3}{3! \langle K^{(3)} \rangle} \sum_{j,k,l=1}^N C_{ijkl} (\delta \psi_j + \delta \psi_k + \delta \psi_l - 3 \delta \psi_i) , \\
	=& \frac{\gamma_3}{2! \langle K^{(3)} \rangle} \sum_{j,k,l=1}^N C_{ijkl}  (\delta \psi_j - \delta \psi_i) ,
	\label{eq:3-simplex_rewritten}
	\end{align}
%
by using the invariance of $C_{ijkl}$ under index permutations, and where the factor $2!=3!/3$ comes from the intermediate step where the expression is written as the sum of 3 identical terms. 

At order 3, definitions \eqref{eq:d-laplacian}-\eqref{eq:d-adj} for the degree $K^{(3)}_{i}$, the adjacency matrix $A^{(3)}_{ij}$, and the Laplacian $L^{(3)}_{ij}$ read
\begin{align}
K^{(3)}_{i} &= \frac{1}{3!} \sum_{j, k,l=1}^N C_{ijkl}  , \\
A^{(3)}_{ij}& = \frac{1}{2!} \sum_{k,l=1}^N C_{ijkl}  , \\ 
L^{(3)}_{ij} &=  3  K^{(3)}_i \delta_{ij}  - A^{(3)}_{ij}   .
\end{align}
We remind the reader that the degree of order 3, $K^{(3)}_{i}$, of node $i$ is the number of distinct 3-simplex interactions it is part of, and $A^{(3)}_{ij}$ is the number of shared 3-simplices including nodes $i$ and $j$.
With these definitions, we can now perform the second important step and rewrite~\eqref{eq:3-simplex_rewritten} as
\begin{align}
\delta \dot \psi_i &=\frac{\gamma_3}{2! \langle K^{(3)} \rangle} \sum_{j,k,l=1}^N C_{ijkl}  (\delta \psi_j - \delta \psi_i) \\
&=  \frac{\gamma_3}{2! \langle K^{(3)} \rangle} \left[ \sum_{j=1}^N 2! A^{(3)}_{ij} \delta \psi_j   -  \delta \psi_i 3! K^{(3)}_{i} \right] , \\
&=  \frac{\gamma_3}{ \langle K^{(3)} \rangle} \sum_{j=1}^N \left[  A^{(3)}_{ij} -  3 K^{(3)}_{i} \delta_{ij} \right] \delta \psi_j , \\
&= -  \frac{\gamma_3}{\langle K^{(3)} \rangle} \sum_{j=1}^N L^{(3)}_{ij} \delta \psi_j .
\end{align}
in a similar way to the usual case of order 1. This shows that the dynamics of the 3-simplex interactions, i.e. of 4 oscillators, can be rewritten in terms of a Laplacian of order 3. Indeed, once again such an operator fulfills the requirements to be a Laplacian, being symmetric, positive semi-definite and zero row-sum.

\subsection{Order $d$}

The rewrite of order $d$ follows the same two important steps as the derivation above. First, we rewrite the term of $d+1$ phases as $d$ terms of 2 phases, and see that they are all equal, yielding
\begin{align}
\delta \dot \psi_i =& + \frac{\gamma_{d}}{{d}! \langle K^{({d})} \rangle} \sum_{j_1, \ldots, j_{{d}}=1}^N \, \, \,  M_{ij_1 \ldots j_{D}} \left( \sum_{m = 1}^{{d}} \delta \psi_{j_m} -  {d} \, \delta \psi_i \right) \\
=& \frac{\gamma_d \, d}{d! \langle K^{(d)} \rangle} \sum_{j_1, \ldots, j_{{d}}=1}^N \, \, \,  M_{ij_1 \ldots j_{D}} (\delta \psi_j -  \, \delta \psi_i) .
\end{align}
With definitions \eqref{eq:d-laplacian}-\eqref{eq:d-adj}, we can now perform the second step, which is to rewrite the this difference of two phases in the terms of the Laplacian of order $d$:
\begin{align}
\delta \dot \psi_i &= \frac{\gamma_d \, d}{d! \langle K^{(d)} \rangle} \sum_{j_1, \ldots, j_{{d}}=1}^N \, \, \,  M_{ij_1 \ldots j_{D}} (\delta \psi_j -  \, \delta \psi_i) \\
&= \frac{\gamma_d}{(d-1)! \langle K^{(d)} \rangle} \left[ \sum_{j=1}^N  2!  A^{(d)}_{ij} \delta \psi_j  -  \delta \psi_i d! K^{(d)}_{i} \right] , \\
&= \frac{\gamma_d}{\langle K^{(d)} \rangle} \sum_{j=1}^N \left[  A^{(d)}_{ij} -  d K^{(d)}_{i} \delta_{ij} \right] \delta \psi_j , \\
&= - \frac{\gamma_d}{\langle K^{(d)} \rangle} \sum_{j=1}^N L^{(d)}_{ij} \delta \psi_j .
\end{align}
This shows that, at any order $d$, the dynamics caused by interactions of order $d$ ($d+1$ oscillators) is determined by the matrix $\lap^{(d)}$. Such a matrix fulfills the properties necessary to be a Laplacian so that its eigenvalues are non-negative and include at least one zero. 

\section{\new{Alternative higher-order coupling functions}}
\label{app:coupling_functions}

As we mentioned in the main text, even restricting ourselves to sine coupling functions that vanish at synchronization, other choices are possible. Here, we show that our framework can readily be used for these other choices. We refer to the coupling functions in system~\eqref{eq:system}, of which there is only one at each order.

\subsection{Order 2}

At order 2, the only other choice is the asymmetric function $\sin(2 \theta_j - \theta_k - \theta_i)$. As in the previous appendix, the first step is to separate the 3-phase term into three 2-phase terms:
\begin{align}
& \frac{\gamma_2}{2!} \sum_{j,k=1}^N B_{ijk} ( 2\delta \psi_j - \delta \psi_k -  \delta \psi_i) , \label{eq:2-simplex} \\
=& \frac{\gamma_2}{2!} \sum_{j,k=1}^N B_{ijk}  (\delta \psi_j - \delta \psi_k) 
+ \frac{\gamma_2}{2!} \sum_{j,k=1}^N B_{ijk}  (\delta \psi_j - \delta \psi_i)  
\end{align}
The second term is as in the symmetric case of the main text, and hence yields $\frac12 L^{(2)}_{ij}$. The first term is different, however, but we can use the symmetry between $j$ and $k$ in this term to show that it vanishes:
\begin{align}
& + \frac{\gamma_2}{2!} \sum_{j,k=1}^N B_{ijk}  (\delta \psi_j - \delta \psi_k)  ,\\
=& + \frac{\gamma_2}{2!} \left[ \sum_{j,k=1}^N B_{ijk}  \delta \psi_j - \sum_{j,k=1}^N B_{ijk}  \delta \psi_k \right] = 0 ,
\end{align}
because $B_{ijk} = B_{ikj}$. Hence, the full term of order 2 with this function is half that with the symmetric function (see Eq.~\eqref{eq:2_laplacian_sys})
\begin{equation}
\delta \dot \psi_i = - \frac{\gamma_2}{\langle K^{(2)} \rangle} \frac12 \sum_{j=1}^N L^{(2)}_{ij} \delta \psi_j .
\end{equation}

Remarkably, this means that, for pure triplets, this choice of coupling function will lead to a convergence that is 2 times slower than the symmetric one.

\subsection{Order $d$}
	
The same two steps can be used to treat any arbitrary $d$, by noticing that, in the first step, each term that includes the phase $i$ yields $1/d$ Laplacian of order $d$, and all other terms vanish as shown for order 2 above. Hence, for oscillators coupled via a general function 
\begin{equation}
\sin(c_1 \theta_{j_1} + c_2 \theta_{j_2} + \dots - c_0 \delta \psi_i) ,
\end{equation}
with integer coefficient $c_j$ such that $\sin(0) = 0$, the linearized dynamics is determined solely by $c_0$ as
\begin{equation}
\delta \dot \psi_i = - \frac{\gamma_d}{\langle K^{(d)} \rangle} \frac{c_0}{d} \sum_{j=1}^N L^{(d)}_{ij} \delta \psi_j . 
\end{equation}

\section{Higher-order degree and adjacency matrices for all-to-all coupling}
\label{app:all-to-all}

In this section, we explicitly write the higher-order adjacency matrix and connectivity defined in section~\ref{subsec:all-to-all-Lap}, for each order $d$, in the higher-order all-to-all setting. 

First, we start with the adjacency matrices of order $d$. In a traditional all-to-all setting, the pairwise adjacency matrix has all entries equal to one, but 0 on the diagonal, which can be written $A_{ij}^{(1)} = 1-\delta_{ij}$. At order $d$, the generic entry $(i,j)$ of the adjacency matrix $A_{ij}^{(d)}$ is equal to the number of distinct $d+1$-oscillator interactions including both oscillators $i$ and $j$, as described above. Hence, the matrix is simply given by the number of ways to pick $d-1$ oscillators among the $N-1$ oscillators left. This number is given by the combinatorics formula $A_{ij}^{(d)}= {N-2 \choose d-1} (1-\delta_{ij})$, where $(1-\delta_{ij})$ ensures that entries with $i=j$ are equal to zero. 

Second, we proceed similarly to write explicitly the degree of order $d$. In a usual all-to-all setting with only pairwise interactions, every oscillators has a pairwise interaction with all $N-1$ oscillators left, i.e. $K_i^{(1)} = N-1$. The degree of order $d$ is equal to the number of ($d+1$)-oscillator interactions that oscillator $i$ is part of. Hence, it is given by the number of ways to pick $d$ oscillators out of the $N-1$ left, which can be written $K_i^{(d)} = {N-1 \choose d}$ in combinatorics notation. \new{Note that it scales with the size of the system as $K_i^{(d)} \sim N^d$.}


\bibliography{bib}

\end{document}